\documentclass[doublecol]{epl2} 

\title{Occurrence of connected clusters in motility-induced phase-separated states of persistent active particles at zero temperature}
\shorttitle{Occurrence of connected clusters of persistent active particles} 

\author{M. Schmiedeberg\inst{1} }
\shortauthor{M. Schmiedeberg}

\institute{                    
  \inst{1} Institut f\"ur Theoretische Physik 1, Friedrich-Alexander-Universität Erlangen-Nürnberg, 91058 Erlangen, Germany
}

\abstract{
  To study the interplay of jamming, cluster formation, and motility-induced phase separation in the zero temperature limit in two dimensions, we consider a simple model system consisting of a bidisperse mixture of disks that are only subject to a repulsion force in case of overlaps and an active force. The orientation of the disks is chosen randomly in the beginning and does not change anytime during the simulation thus corresponding to an infinite persistence length.
Simulations with our model system reveal that jammed clusters of particles occur in the dense phase of a phase-separated state in case of intermediate values of the ratio of active to repulsive force. However, for smaller activities there are only a few overlaps between the particles in the dense phase and the coexistence region ends at a packing fraction below the onset of jamming. Finally, for large activities small clusters corresponding to small patches of the dense phase are found that are unstable due to the activity. Our findings on how jamming and phase separation are related are relevant to many active particle systems whose zero temperature and long persistence limits correspond to our model system. 
}

\begin{document}

\maketitle


\section{Introduction}

Active Brownian particles, i.e., particles that use energy from their surrounding to self-propel, are widely-studied as model system in non-equilibrium statistical physics (for reviews see, e.g., \cite{Ramaswamy,vicsek,Marchetti,Aranson,Bechinger}). A phenomenon of special interest is the formation of clusters or active crystals that has been reported in simulations \cite{marchetti,redner,fily,siebert,dig,lee,nie}, in theoretical approaches \cite{bialke13,cates13,Speck,Bialke,Schmidt}, and experiment \cite{Buttinoni,Palacci,geyer}. Usually the cluster formation is related to a non-equilibrium phase separation termed motility-induced phase separation (for a review see \cite{cates}) where the system splits into a gas-like dilute phase and dense phase. Typically the packing fraction and the P{\'e}ctlet number which is given by the ratio of the strength of activity and the diffusion constant are used as control parameters.

The relation of motility-induced phase separation and jamming or glassy dynamics in systems with large P{\'e}ctlet numbers corresponding to a large persistence length has been studied in recent works \cite{ciamarra,berthier}. Dynamically jammed states or glass-like states have been reported to occur for densities above the coexistence region \cite{ciamarra,berthier}.

Here we study the case of an infinite P{\'e}ctlet number. However, as an additional control parameter we vary the ratio of the active force and the repulsion force, which is small in \cite{ciamarra,berthier}. We are not interested in glassy dynamics but in jamming defined by overlaps as in passive systems.

In passive athermal systems jamming can be studied by starting with a completely random system (corresponding to infinite temperature), minimizing the energy without crossing energy barriers (corresponding to a quench to zero temperature), and analyzing whether overlaps remain in the final configuration, which then is called jammed, or whether there are no overlaps denoting an unjammed state \cite{Ohern,Ohern2}. In large systems, the number of overlaps per particle jump from zero below the packing fraction $\phi_J=0.842$ of the jamming transition to $4$ in two dimensions \cite{Ohern}. This kind of jamming corresponds to a transition from a state without overlaps to a state where almost all particles (except for a few ones that are termed rattlers \cite{Ohern2}) are part of a cluster where all particles are connected to other particles in the cluster via overlaps. As in an active phase-separated state not all particles are in the same phase and usually only some of the particles are jammed in a part of the system, we here search for large connected clusters. Furthermore, we always use the term jamming in such a structural sense but not in the dynamical sense that has been considered in other works \cite{henkes,ciamarra}. 

In a previous work \cite{maiti2019} we have combined the energy minimization as in \cite{Ohern,Ohern2} with an active force along a direction that is initially randomly assigned to each particle and that does not change during the simulation. Overlap-free final states correspond to unjammed states while states with remaining overlaps correspond to states with jammed clusters. For a three dimensional system we have found a discontinuous transition in case of small activity occurring at packing fractions below $0.55$ and a continuous transition for larger activities at packing fractions between $0.55$ and $0.64$ \cite{maiti2019} where the value $0.64$ corresponds to the value of the packing fraction of the athermal jamming transition in a passive system in three dimensions \cite{Ohern,Ohern2}. While the energy minimization protocols used in \cite{Ohern,Ohern2,maiti2019} are not related to any dynamics, we here want to study the occurrence of jammed clusters (now in two dimensions) with simulations based on real dynamics.

\section{Model}

We consider a bidisperse mixture of $N$ disks in two dimensions. Half of the particles possess a diameter $\sigma$ while the diameter of the other disks is $1.4\sigma$. Due to the bidispersity crystallization is suppressed.

Brownian dynamics simulations at zero temperature are employed. Therefore, an overdamped Langevin-like equation without thermal forces is integrated, i.e., 
\begin{equation}
\label{eq.1}
\gamma \dot{\vec{r}}=\vec{F}_r+\vec{F}_a
\end{equation}
where $\gamma$ is the friction constant, $\vec{F}_{r}$ denotes the interaction forces and $\vec{F}_a$ a force denoting the activity of the particles. Specifically, for the interactions a constant repulsion force of magnitude $F_r$ acts for each overlap, while there is no interaction without overlap. The force on a particle due to activity is $\vec{F}_a=F_a\vec{\theta}$ where $\vec{\theta}$ is the orientation of the particle.

At the beginning a bidisperse mixture of $N$ particles are randomly placed in a box with periodic boundary conditions. The box size $A_t=L^2$ is chosen such that a given packing fraction $\phi=N\pi\sigma^2(1+1.4^2)/(8L^2)$ is obtained. For each particle an individual random orientation $\vec{\theta}$ is chosen at the beginning of the simulation that does not change at any time, i.e., there is no rotational diffusion as an athermal system is considered. As a consequence, the persistence length and the P{\'e}ctlet number of the motion are both infinite.

The control parameters that we vary in our study are the packing fraction $\phi$ and the ratio $F_a/F_r$, i.e., the magnitude of the active forces divided by the strength of the repulsion.

Usually we consider $N=20000$ particles that are simulated at least until a time $100000 F_a/\gamma$ where a steady state is reached. We performed some test runs with $N=10000$ or $N=40000$ and did not observe a different behavior. Furthermore, simulation runs initialized in a phase-separated state or simulations running for much longer times did not lead to different results. Note that for a passive system ($F_a/F_r=0$) we can observe the athermal jamming transition at $\phi_{\textnormal{J}}=0.84$ in agreement with \cite{Ohern}.

\section{Results}

\begin{figure*}
\onefigure[width=\linewidth]{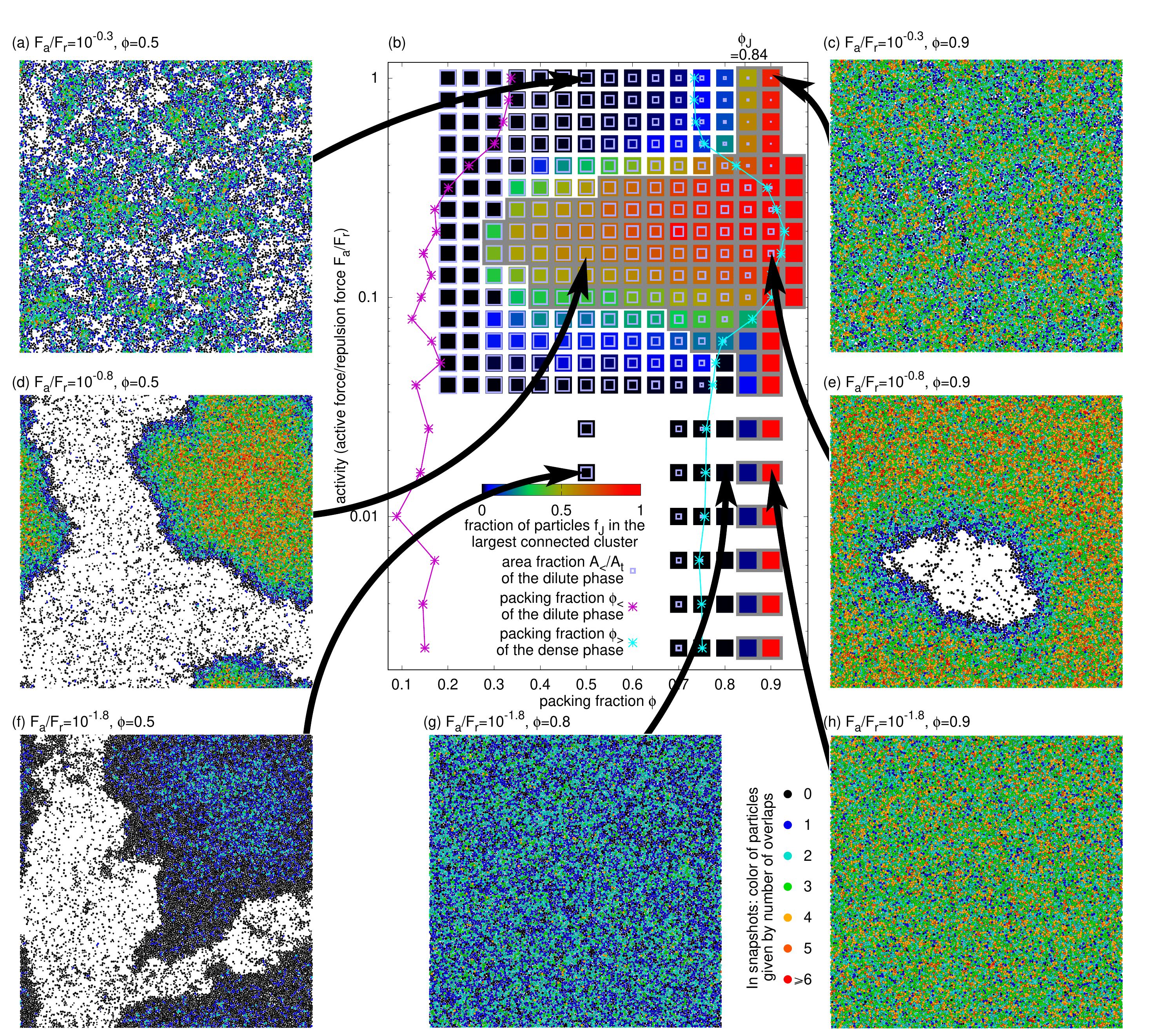}
\caption{(a,c-h)} Snapshots of the active particles in the steady state. The disks are colored according to the number of overlapping neighboring particles as specified in the legend between panel (g) and (h). (b) State diagram as function of the packing fraction $\phi$ and the ratio $F_a/F_r$ of the magnitude of the active forces and the strength of the repulsion. The colors denote the fraction of particles $f_J$ that are in the largest cluster that is formed of disks in contact. The open squares in light-gray mark the area fraction $A_</A_t$ of the dilute phase with a large square indicating that almost all particles are in the dilute state while small squares mean that almost all particles are in the dense state. Note that $A_</A_t$ for the large density cases where the dilute phase starts to vanish is shown in more detail in Fig.\ \ref{fig.2}(b). In case the dense phase possesses a local packing fraction larger or equal to the athermal jamming density $\phi_{\textnormal{J}}=0.84$ a gray background is drawn. Finally, the packing fractions $\phi_<$ of the dilute and $\phi_>$ the dense phase for phase-separated states at $\phi=0.7$ are shown by stars and lines in magenta and cyan, respectively. The packing fractions at other overall $\phi$ are plotted in Fig.\ \ref{fig.2}(a).
\label{fig.1}
\end{figure*}

As expected, we observe a separation into a dilute and a dense phase for intermediate packing fractions. In the dilute phase only a few particles (usally without contacts) can be found while in the dense phase many particles are overlapping with neighboring disk or at least are close to them. Typical examples are shown as snapshots in Figs.\ \ref{fig.1}(d,e,f).

Fig.\ \ref{fig.1}(b) is a state diagram, where as a function of $\phi$ and $F_a/F_r$ we show what fraction of the area is occupied by the dilute phase (indicated by the size of the open light-gray squares), how many particles are in the largest connected cluster (denoted by the colors), and whether the local packing fraction of the dense phase is above the packing fraction $\phi_J=0.84$ where in equilibrium athermal jamming in two dimensions occurs \cite{Ohern} (marked by a gray background). Furthermore, the magenta and cyan stars indicate the packing fractions within the two phases obtained in the coexistence region at $\phi=0.7$. In the following, we will first explain all these quantities in detail, before we discuss the results.

To quantify the phase separation, we first calculate the Voronoi cells for all particles. Based on their areas, we determine a local packing fraction for each particle. The sum of the areas of the Voronoi cells with local packing fractions below $0.4$ is termed the area of the dilute phase $A_<$. In Fig.\ \ref{fig.1}(b) the fraction $A_</A_t$, where $A_t$ is the total area of the simulation box, is indicated by open squares ranging from large squares in case almost the whole system is in the dilute phase to small squares for the case where almost all disks are in the dense phase. Note that the threshold packing fraction of $0.4$ was chosen because it is the middle between the observed dilute gas-like phase and the jammed or almost jammed dense phase. Furthermore, for phase-separated states there are only very few particles that have a local packing fraction around $0.4$. 

By averaging all local packing fractions that are smaller than $0.4$ and all values that are larger than $0.4$, we obtain the the packing fractions $\phi_<$ and $\phi_>$ of the dilute and the dense phase, respectively. In Fig.\ \ref{fig.1}(b) $\phi_<$ and $\phi_>$ as obtained at an overall packing fraction $\phi=0.7$ are plotted with magenta and cyan stars, respectively. For phase separation scenarios in equilibrium the densities of the phases in coexistence usually are constant when scanning through the coexistence region by varying the overall density. Therefore, in such a case $\phi_<$ and $\phi_>$ obtained at any packing fraction within the coexistence region would give the packing fraction where the coexistence region starts and the packing fraction where the coexistence region ends. In our non-equilibrium system, $\phi_>$ roughly marks the packing fractions where the dilute phase disappears except for large $F_a/F_r$. To be specific, for $F_a/F_r<0.4$, the packing fractions of the stars in cyan in Fig.\ \ref{fig.1}(b) coincide with the packing fractions where $A_</A_t$ becomes zero. However, as we will discuss later, the local packing fractions within the coexistence region is not constant if the overall packing fraction is changed.

We are also interested to determine whether there is a significant number of overlaps between the disks in the dense phases such that it might be compared to a state that is called jammed in a passive system. In the case shown in the snapshots in Fig.\ \ref{fig.1}(d) that is typical for intermediate activities with $F_a/F_r$ between $0.09$ and $0.35$ most of the disks in the bulk of the dense phase possesses three or more neighboring particles in contact. In Fig.\ \ref{fig.1}(f) that is typical for smaller $F_a/F_r$ a similar phase separation occurs. However, hardly any particle has more than two contacts. Interestingly, the contacts in Fig.\ \ref{fig.1}(f) are not sufficient to form a larger connected cluster. To be specific, in Fig.\ \ref{fig.1}(f), the largest cluster where all particles are connected to each other along overlaps contains 15 particles while in Fig.\ \ref{fig.1}(d) 12882 of the 20000 particles are connected to each other via contacts. In a passive athermal system after energy minimization, the particles within the large clusters would be called jammed as overlaps are present. In that sense, the states with intermediate $F_a/F_r$ might be called jammed while the states with small $F_a/F_r$ only posses a few small jammed clusters but otherwise are unjammed. 

To quantify in what cases a significant number of particles is part of a larger connected cluster, we first determine all clusters where all disks are connected by overlaps to other particles of the cluster. Then we search for the largest of these clusters and calculate what fraction of particles $f_J$ is part of this cluster. The result is indicated by the colors in Fig.\ \ref{fig.1}(b). For $F_a/F_r$ between $0.09$ and $0.35$ and packing fractions above $0.3$ the majority of the particles can be part of a connected cluster. In contrast, for larger or smaller activities a large connected cluster only occurs in case the packing fraction is increased above $\phi_J=0.84$.

\begin{figure}
\onefigure[width=\linewidth]{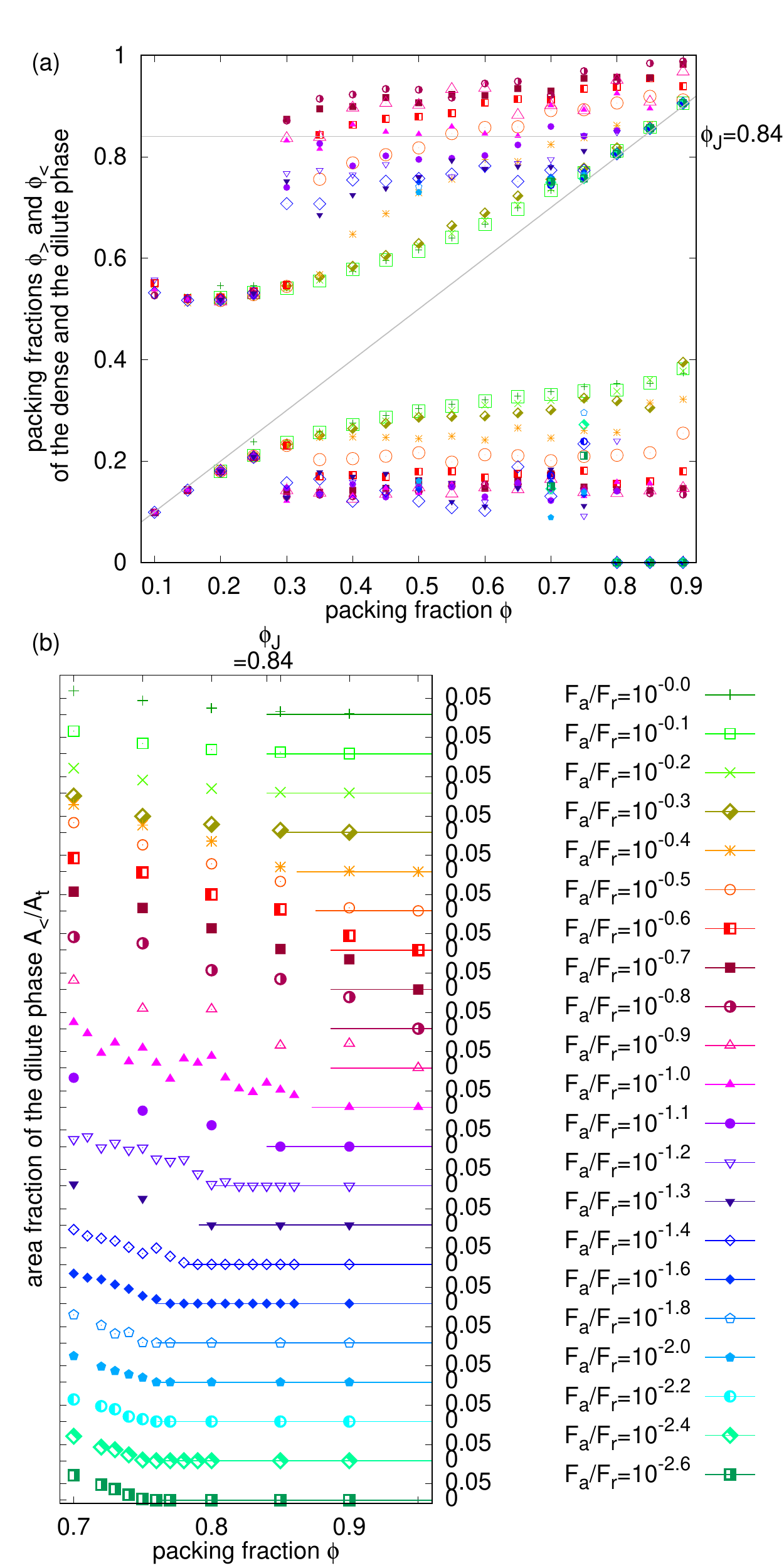}
\caption{(a) Packing fractions $\phi_<$ and $\phi_>$ as a function of the overall packing fraction $\phi$ of a system. The same symbols as indicated in the legend of (b) are used. (b) Fraction of the area of the dilute phase $A_</A_t$ at large packing fractions $\phi$ for various activities. The colored horizontal lines indicate $A_</A_t=0$.}
\label{fig.2}
\end{figure}

As mentioned before, in our non-equilibrium system the packing fractions $\phi_<$ and $\phi_>$ in the phase-separated states might depend on the overall packing fraction $\phi$. Therefore, we plot $\phi_<$ and $\phi_>$ as a function of $\phi$ in Fig.\ \ref{fig.2}(a). For large $F_a/F_r$ (see, e.g., the green squares for $F_a/F_r=10^{-0.1}$), $\phi_<$ and $\phi_>$ are continuous in $\phi$. However, if $F_a/F_r$ is decreased a jump around $\phi=0.3$ occurs. For example, the red open circles for $F_a/F_r=10^{-0.5}$ jump between $\phi=0.3$ and $\phi=0.35$ and $\phi_>$ further increases afterwards and even exceeds $\phi_J=0.84$ for $\phi\geq 0.55$. For $F_a/F_r=10^{-0.9}$ (magenta open upward triangles) the jump occurs between $\phi=0.25$ and $\phi=0.3$ and the local packing fraction $\phi_>$ is close to $\phi_J$ right after the jump. However, if $F_a/F_r$ is decreased further, the jump becomes smaller again and $\phi_J$ is not reached before the overall density $\phi$ exceeds $\phi_J$ (see, e.g., blue diamonds for $F_a/F_r=10^{-1.4}$). In Fig.\ \ref{fig.1}(b) the cases where $\phi_>\geq \phi_J$ are marked by a gray background. These cases coincide with the systems where most particles are in the largest connected cluster.

The local packing fraction $\phi_>$ possesses a maximum for $F_a/F_r\approx 0.2$ while it is smaller for both smaller and larger values of $F_a/F_r$. Accordingly, the packing fraction where the coexistence region ends and only a homogeneous dense phase is observed is larger for intermediate $F_a/F_r$ than for small $F_a/F_r$. In Fig.\ \ref{fig.2}(b) we take a closer look on the area fraction $A_</A_t$ of the dilute phase as a function of $\phi$ for various $F_a/F_r$. For large $F_a/F_r\geq 10^{-0.2}$ the area fraction $A_</A_t$ becomes small at $\phi_J$ but strictly speaking is still nonzero for larger $\phi$ as the large activity leads to small dilute voids even at large $\phi$. For $10^{-1.0}\leq F_a/F_r\leq 10^{-0.3}$ the area fraction $A_</A_t$ is still larger than zero at packing fractions around $\phi_J=0.84$ but vanishes between $\phi=0.9$ and $\phi=0.95$. For smaller $F_a/F_r$ the area fraction $A_</A_t$ goes to zero at packing fractions that are significantly smaller than $\phi_J=0.84$. For the smallest considered values of $F_a/F_r$ the packing fraction where $A_</A_t$ vanishes is $\phi=0.76\pm 0.01$.

\section{Summary and Discussion}

Our results demonstrate that motility-induced phase separation and the occurrence of connected clusters are linked to each other and depend on the ratio $F_a/F_r$ of active to repulsive force. We find three different regimes:

For large $F_a/F_r$ dilute and dense patches are visible (see, e.g., Fig.\ \ref{fig.1}(a)). However, due to the strong activity no larger dense clusters or dilute regions occur. As a consequence no large connected clusters can be formed, except at packing fractions above $\phi_J$. Note that even in the jammed structures at large packing fractions small bubbles of the dilute phase can be found (see, e.g., Fig.\ \ref{fig.1}(c)).

For intermediate $F_a/F_r$ the packing fraction $\phi_>$ of the dense phase possesses a jump as a function of $\phi$. After the jump the local packing fraction reach values above $\phi_J$ and accordingly a significant number of overlaps is observed in the dense phase leading to large connected clusters. The dilute phase is still present at $\phi=\phi_J$ and only disappears at much larger packing fractions between $0.9$ and $0.95$.

For small $F_a/F_r$ large clusters of connected particles do not occur in the coexistence region. The packing fraction $\phi_>$ of the dense phase stays below $\phi_J$. Furthermore, the phase separation ends at a packing fraction below $\phi_J$. The systems considered in \cite{ciamarra,berthier} probably best compares to our small $F_a/F_r$ case.

Our results indicate that for intermediate $F_a/F_r$ the repulsive force is too weak to efficiently remove overlaps. Therefore, a significant number of overlaps remains, the packing fraction is larger than for small $F_a/F_r$ and as it even exceeds $\phi_J$ in the dense phase, large connected clusters occur corresponding to what in a passive system is called a jamming transition.

\section{Conclusions and Outlook}

We show that the onset of jamming might occur in phase-separated states in some cases while it is only found for packing fractions above the coexistence region in other cases. The competition between activity, which usually creates new overlaps, and repulsion forces that try to remove overlaps is crucial to understand the interplay of jamming and phase separation.

There are obvious extensions that might be interesting for future works. One question is whether all cases that we report can still be observed in a system with finite temperature that leads to translational as well as rotational thermal motion and thus to a finite persistence length. Note that there are some works that report that glassy dynamics sets in at packing fractions above the coexistence region \cite{ciamarra,berthier}. Probably these systems correspond to our cases with small $F_a/F_r$.

Furthermore, the type of transition and its critical behavior for jamming certainly is of great interest. Note that unlike in a passive athermal system \cite{Ohern,Ohern2} or the system with an artificial energy minimization \cite{maiti2019}, we never get rid of all overlaps and thus with the system studied here we can never observe a transition that is as clear as in \cite{Ohern,Ohern2,maiti2019}. According to \cite{maiti2019} a discontinuous transition is expected for intermediate $F_a/F_r$, while the transition for small $F_a/F_r$ is continuous. Note that we observe a jump in the packing fractions $\phi_<$ and $\phi_>$, which might be seen as indication of a discontinuous transition. Whether the observed continuous increase of the size of the largest jammed cluster in the small $F_a/F_r$ is an indication of an continuous transition or is the consequence of our protocol requires further work. Note that a continuous jamming transition would  indicate a singular limit for the case $F_a/F_r\rightarrow 0$ as the athermal jamming transition in a passive system with $F_a/F_r$ is discontinuous \cite{Ohern,Ohern2}. Note there are also other perturbations or variations of the discontinuous athermal jamming transition that lead to a continuous transition. For example, modification of random organization protocols that describe a continuous transition can approach athermal jamming \cite{milz,chaikin}. Furthermore, random steps during the energy minimization protocol of athermal jamming might lead to a continuous transition that is in the same universality class as random organization \cite{maiti2018,maiti2018c}. Therefore, if activity is considered as another perturbation of the athermal jamming transition maybe the consequences are similar as for the perturbations studied in \cite{maiti2019,chaikin,maiti2018,maiti2018c,corwin17}.

\acknowledgments
I want to thank M. Maiti, E. O\u{g}uz, and A. Menzel for helpful discussions.


\begin{thebibliography}{0}



\bibitem{Ramaswamy}
\Name{Ramaswamy S.}
   \REVIEW{Annu. Rev. Condens. Matter Phys.} {1} {2010} {323}


\bibitem{vicsek}
  \Name{Vicsek T. \and Zafeiris A.}
  \REVIEW{Phys. Rep.}{71}{2012}{140}
  
\bibitem{Marchetti}
  \Name{Marchetti M. C., Joanny J. F., Ramaswamy S., Liverpool T. B., Prost J., Rao M. \and Simha R. A.}
  \REVIEW{Rev. Mod. Phys.} {85} {2013} {1143}

\bibitem{Aranson}
  \Name{Aranson I.}
  \REVIEW{Comptes Rendus Physique}{14} {2013} {518}
  
\bibitem{Bechinger}
  \Name{Bechinger C, Di Leonardo R., L\"owen H., Reichhardt C., Volpe G. \and Volpe G.}
  \REVIEW{Rev. Mod. Phys.} {88} {2016} {045006}


\bibitem{marchetti}
\Name{Fily Y. \ and Marchetti M.C.}
\REVIEW{Phys. Rev. Lett.} {108} {2012} {235702}

  \bibitem{redner}
    \Name{Redner G.S., Hagan M.F. \and Baskaran A.}
    \REVIEW{Phys. Rev. Lett.}{110}{2013} {055701}
    
  \bibitem{fily}
    \Name{Fily Y., Henkes S. \and Marchetti M.C.}
    \REVIEW{Soft Matter}{10}{2014} {2132}

  \bibitem{siebert}
    \Name{Siebert J.T., Dittrich F., Schmid F., Binder K., Speck T. \and Virnau P.}
    \REVIEW{Phys. Rev. E}{98}{2018} {030601(R)}

  \bibitem{dig}
    \Name{Digregorio P., Levis D., Suma A., Cugliandolo L.F., Gonnella G. \and Pagonabarraga I.}
    \REVIEW{Phys. Rev. Lett.}{121}{2018} {098003}

  \bibitem{lee}
    \Name{Partridge B. \and Lee C.F.}
    \REVIEW{Phys. Rev. Lett.}{123}{2019} {068002}
    
  \bibitem{nie}
    \Name{Nie P., Chattoraj J., Piscitelli A., Doyle P., Ni R. \and Pica Ciamarra M.}
    \REVIEW{Phys. Rev. Res.}{2}{2020} {023010}


    

  \bibitem{bialke13}
    \Name{Bialk{\'e} J., L\"owen H. \and Speck T.}
    \REVIEW{EPL}{103}{2013} {30008}
    
  \bibitem{cates13}
    \Name{Stenhammar J., Tiribocchi A., Allen R.J., Marenduzzo D.
\and Cates M.E.}
    \REVIEW{Phys. Rev. Lett.}{111}{2013} {145702}

  \bibitem{Speck}
    \Name{Speck T., Bialk\'e J., Menzel A. M. \and L\"owen H.}
    \REVIEW{Phys. Rev. Lett.}{112}{2014} {218304}
    
 \bibitem{Bialke}
  \Name{Bialk\'e J., Speck T. \and L\"owen H.}
  \REVIEW{Journal of Non-Crystalline Solids}{407}{2015}{367}

\bibitem{Schmidt}
    \Name{Hermann S., Krinninger P., de las Heras D. \and Schmidt M.}
    \REVIEW{Phys. Rev. E}{100}{2019} {052604}



\bibitem{Buttinoni}
    \Name{Buttinoni I., Bialk\'e J., K\"ummel F., L\"owen H., Bechinger C. \and Speck T.}
    \REVIEW{Phys. Rev. Lett.} {110}{2013} {238301}

 \bibitem{Palacci}
   \Name{Palacci J., Sacanna S., Steinberg A. P., Pine D. J. \and Chaikin P. M.}    \REVIEW{Science} {339}{2013} {936}

\bibitem{geyer}
    \Name{Geyer D., Martin D., Tailleur J. \and Bartolo D.}
    \REVIEW{Phys. Rev. X} {9}{2019} {031043}



  
\bibitem{cates}
\Name{Cates M.E. \and Tailleur J.}
   \REVIEW{Annu. Rev. Condens. Matter Phys.} {6} {2015} {219}


   \bibitem{ciamarra}
\Name{Yang J., Ni R. \and Pica Ciamarra M.}
   \REVIEW{Phys. Rev. E} {106} {2022} {L012601}

   \bibitem{berthier}
\Name{Keta Y.-E., Robert L. Jack R.L. \and Berthier L.}
   \REVIEW{Phys. Rev. Lett.} {129} {2022} {048002}
 

 \bibitem{Ohern}
\Name{O'Hern C.S., Langer S.A., Liu A.J. \and Nagel S.R.}
   \REVIEW{Phys. Rev. Lett.} {88} {2002} {075507}

\bibitem{Ohern2}
\Name{O'Hern C.S., Silbert L.E., Liu A.J. \and Nagel S.R.}
\REVIEW{Phys. Rev. E}  {68} {2003} {011306}

\bibitem{henkes}
\Name{Henkes S., Fily Y., and Marchetti M.C.}
   \REVIEW{Phys. Rev. E} {84} {2011} {040301(R)}
   


\bibitem{maiti2019}
\Name{Maiti M. \and Schmiedeberg M.}
   \REVIEW{EPL}{126}{2019}{46002}



\bibitem{milz}
\Name{Milz L. \and Schmiedeberg M.}
   \REVIEW{Phys. Rev. E}{88}{2013}{062308}
   
\bibitem{chaikin}
\Name{Sam Wilken S., Guerra R.E., Levine D. \and Chaikin P.M.}
   \REVIEW{Phys. Rev. Lett.}{127}{2021}{038002}
   
  
\bibitem{maiti2018}
\Name{Maiti M. \and Schmiedeberg M.}
   \REVIEW{Scientific Reports} {8} {2018} {1837}

\bibitem{maiti2018c}
\Name{Maiti M. \and Schmiedeberg M.}
   \REVIEW{Eur. Phys. J. E}{42}{2019}{38}

\bibitem{corwin17}
   \Name{Morse, P.K. \and Corwin, E.I.}
\REVIEW{Phys. Rev. Lett.}{119}{2017}{118003}

\end{thebibliography}
\end{document}